\documentclass[10pt,letterpaper,twocolumn]{article} 

\usepackage{ol2}
\usepackage[draft]{hyperref}
\usepackage{amsmath}
\begin{document}

\twocolumn[ 

\title{Rectification of light refraction in curved waveguide arrays}


\author{Stefano Longhi}
\address{Dipartimento di Fisica and Istituto di Fotonica e
Nanotecnologie del CNR \\ Politecnico di Milano, Piazza L. da
Vinci, 32 I-20133 Milano (Italy)
\\}

\begin{abstract}
An 'optical ratchet' for discretized light in photonic lattices,
which enables to observe rectification of light refraction at any
input beam conditions, is theoretically presented, and a possible
experimental implementation based on periodically-curved zigzag
waveguide arrays is proposed.
\end{abstract}



] 

\noindent Transport of discretized light in photonic lattices has
attracted a great attention in recent years from both fundamental
and applied viewpoints \cite{Christodoulides03,Lederer08}.
Engineered photonic lattices offer the possibility of tailoring
diffraction \cite{Eisenberg00}, refraction \cite{Pertsch02}, and
of shaping polychromatic light beams
\cite{Garanovich06,Neshev07,Sukhorukov07}. They also provide a
rich laboratory system to mimic the dynamics of quantum particles
driven or disordered periodic lattices, with the observation of
the optical analogues of Bloch oscillations
\cite{Christodoulides03,Trompeter06}, Anderson localization
\cite{Segev07,Lahini08} and dynamic localization
\cite{Longhi06,Aitchison07,Dreisow08} to name a few. Recently, the
transport of ac-driven quantum particles in periodic lattices has
received a great interest also for the possibility to achieve
Hamiltonian ratchet effects, i.e. a rectified transport in absence
of a net bias force and dissipation. Different ratchet models
based on optical lattices, which use ultracold atoms as ideal
systems, have been proposed
\cite{Gong07,Denisov07a,Creffield07,Denisov07b,Isart07,Molina08},
and detailed studies have investigated the relation between
rectification and broken space-time symmetries
\cite{Denisov07a,Denisov07b}, the effects of atomic interactions
\cite{Molina08}, and the possibility to transport quantum
information \cite{Creffield07,Isart07}. In optics, the idea of
realizing an 'optical ratchet' has received few attention to date,
and limited to solitary waves in nonlinear systems
\cite{Gorbach06,Kartashov05,Kartashov06}. In particular, in
\cite{Gorbach06} a setup to observe ratchet effects for cavity
solitons has been proposed using coupled waveguide optical
resonators with an adiabatically-shaken holding pump beam, whereas
in \cite{Kartashov05} control of soliton dragging in dynamic
optical lattices has been investigated.\\
In this Letter a simple ratchet scheme is proposed in a {\it
linear} optical system corresponding to rectification of
discretized light refraction in an array of periodically-curved
zigzag waveguides, which may be realized with currently available
femtosecond laser writing technology \cite{Dreisow08b}. We
consider two interleaved arrays of single-mode channel waveguides,
a primary array  A$_n$ and an auxiliary array B$_n$, in the zigzag
geometry shown in Fig.1(a), which is similar to the one recently
realized in Ref.\cite{Dreisow08b} to investigate second-order
coupling effects. As opposed to Ref.\cite{Dreisow08b}, the
propagation constant of modes in waveguides B$_n$ is assumed to be
detuned by a relatively large amount $\sigma$ from the propagation
constant of modes in waveguides A$_n$, a condition that can be
achieved in practice by e.g. lowering the writing velocity of the
auxiliary array. The optical axis of the waveguides, which lies in
the $(X,Z)$ plane, is assumed to be periodically-curved along the
paraxial propagation direction $Z$ with a bending profile $x_0(Z)$
of period $2z_0$. In the waveguide reference frame $z=Z$ and
$x=X-x_0(Z)$, where the arrays appear to be straight, under scalar
and paraxial assumptions light propagation at wavelength $\lambda$
is described by the following Schr\"{o}dinger-like equation (see,
for instance, \cite{Longhi06})
\begin{equation}
i \hbar \frac{\partial \psi}{\partial z}=-\frac{\hbar^2}{2 n_s}
\nabla^2 \psi+[n_s-n(x,y)] \psi+n_s \ddot{x}_{0}(z) x \psi
\end{equation}
where $\hbar= \lambda/(2 \pi)$ is the reduced wavelength, $n_s$ is
the refractive index of the substrate medium, $n(x,y)$ is the
refractive index profile of the (straight) arrayed structure,
$\nabla^2$ is the transverse Laplacian, and the dot denotes the
derivative with respect to $z$. As discussed e.g. in
\cite{Longhi06}, light propagation in the optical system mimics
the temporal dynamics of a quantum particle in a binding potential
$n_s-n(x,y)$ driven by an external (inertial) force
$-n_s\ddot{x}_0(z)$ with zero mean. In the tight-binding
approximation, light transfer among the waveguides may be
described by a set of coupled mode equations for the amplitudes
$a_n$ and $b_n$ of modes trapped in waveguides A$_n$ and B$_n$,
respectively, which may be cast in the following form (see, for
instance, \cite{Garanovich08})
\begin{eqnarray}
i \dot{a}_n & = &   -\kappa_{aa} \left[ a_{n+1} \exp(-2 i
\Theta)+a_{n-1} \exp( 2 i\Theta) \right] +  \\
 & - &   \kappa_{ab}
\left[  b_n \exp( i \sigma z - i \Theta) +b_{n-1} \exp( i \sigma z + i \Theta) \right], \nonumber \\
i \dot{b}_n & = &   -\kappa_{bb} \left[ b_{n+1} \exp(-2i
\Theta)+b_{n-1} \exp( 2i\Theta) \right] + \\
 & - &   \kappa_{ab}
\left[  a_n \exp(- i \sigma z + i \Theta) +a_{n+1} \exp( -i \sigma
z - i \Theta) \right], \nonumber
\end{eqnarray}

\begin{figure}[htb]
\centerline{\includegraphics[width=8.2cm]{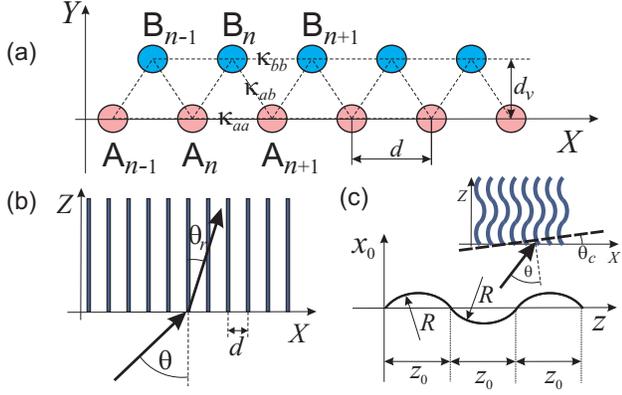}} \caption{
(Color online) (a) Schematic of a zigzag waveguide array, composed
by a primary array A$_n$ and an auxiliary array B$_n$. (b) Light
refraction in a straight waveguide array. (c) Axis bending profile
$x_0(z)$ for the observation of light refraction rectification.
The inset in (c) shows a schematic of the periodically-curved
array.}
\end{figure}

\begin{figure}[htb]
\centerline{\includegraphics[width=8.2cm]{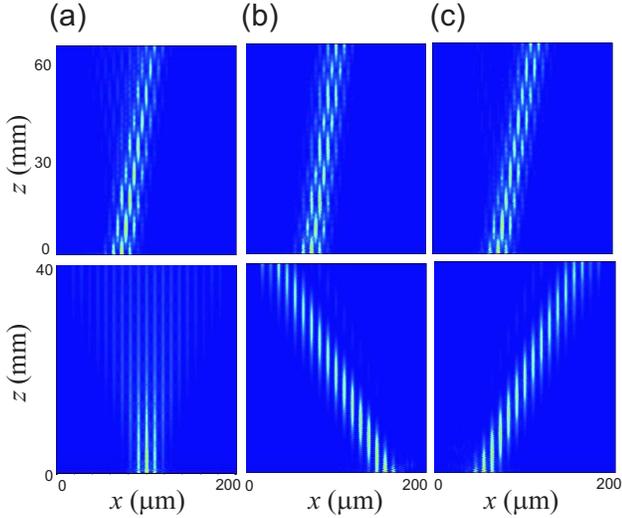}}
\caption{(Color online) Rectified refraction (pseudocolor map
 of integrated intensity distribution) in a 65-mm-long periodically-curved zigzag waveguide array (upper images),
 and corresponding behavior expected for a straight zigzag array (bottom images), for an input elliptical beam exciting a few waveguides in the primary
 array at different
 tilting angles: (a) $\theta=0$, (b) $\theta=-0.5 \theta_B$, and (c) $\theta=0.5 \theta_B$. The values of other parameters are: $\lambda=633$ nm,
 $n_s=1.45$, $\Delta n_A=0.003$, $\Delta n_B=0.0032$, $w=3 \; \mu$m, and $d=9 \; \mu$m.}
\end{figure}

In Eqs.(2) and (3), $\kappa_{aa}$ and $\kappa_{bb}$ are the
coupling rates between adjacent waveguides in the primary and
auxiliary arrays, respectively, $\kappa_{ab}$ is the
cross-coupling rate [see Fig.1(a)], $\Theta(z)= \pi n_s
\dot{x}_0(z)d/ \lambda$, $d$ is the spatial period of the arrays,
and $\sigma$ is the propagation
constant mismatch between waveguides in the two arrays.\\
 For straight arrays and large mismatch
$|\sigma| \gg \kappa_{ab},\kappa_{aa},\kappa_{bb}$, light dynamics
in the two arrays is decoupled and a broad beam, which excites the
primary array A$_n$ at an incidence angle $\theta$ smaller than
the Bragg angle $\theta_B=\lambda/(2d)$ [see Fig.1(b)], will
diffract and refract at an angle $ \theta_r(\theta)=2 \kappa_{aa}d
\sin( \pi \theta / \theta_B)$ as in an ordinary tight-binding
linear lattice \cite{Pertsch02}. Rectification of beam refraction,
i.e. a net transverse drift of discretized beams averaged over all
possible incidence angles $\theta$, is obviously absent because
$\theta_r(-\theta)=-\theta_r(\theta)$. Even for a single array
A$_n$, curved to mimic a dc, ac or combined dc-ac driving fields,
a net transport of wave packets -averaged over different initial
conditions- is generally unlikely, as discussed in
Ref.\cite{Gong07}. To achieve a ratchet effect, we introduce the
auxiliary lattice B$_n$ and assume that both arrays are
periodically-curved with a bending profile $x_0(z)$ made of a
sequence of circular arcs of (paraxial) length $z_0$ and constant
curvature with alternating sign $\ddot{x}_0(z)= \pm 1/R$, as
depicted in Fig.1(c). Such a bending profile corresponds to an ac
square-wave force $-n_s \ddot{x}_0(z)$, and to a triangular shape
for the function $\Theta(z)$ entering in Eqs.(2) and (3). Note
that, since in this case the waveguide axis at the entrance plane
$z=0$ is tilted by an angle $\dot{x}_0(0)=z_0/(2R)$, the incidence
angle $\theta$ should be now measured with respect to a plane
tilted by an angle $\theta_c=n_s z_0/(2R)$ from the entrance facet
\cite{Longhi06}, as shown in the inset of Fig.1(c). To achieve
rectification of light refraction in the primary array over one
undulation cycle, the radius of curvature $R$ and spatial period
$z_0$ of alternations are designed to satisfy the conditions
$\dot{\Theta}(z)= \pm \sigma$ and $\kappa_{ab} z_0= \pi/2$, i.e.
\begin{equation}
R=\pi n_s d/(\lambda |\sigma|), \; \; z_0=\pi/(2 \kappa_{ab}).
\end{equation}
In the first semi-cycle of undulation, where $\dot{\Theta}(z)=
\sigma$, neglecting nonresonant terms Eqs.(2) and (3) reduce to
$i\dot{a}_n=-\kappa_{eff}b_n$ and $i\dot{b}_n=-\kappa_{eff}^{*}
a_n$, where $\kappa_{eff}=\kappa_{ab} \exp(i \phi)$ and
$\phi=\sigma z- \Theta$ is a constant phase shift. This means that
waveguide A$_{n}$ turns out to be coupled solely with waveguide
B$_{n}$, and light transfer between them occurs like in an
ordinary synchronous directional coupler. For a propagation length
$z_0$ equal to one coupling length $z_0=\pi/(2 \kappa_{ab})$,
light trapped in waveguide A$_n$ is thus fully transferred into
waveguide B$_n$. At the successive propagation semi-cycle, where
$\dot{\Theta}(z)= -\sigma$, Eqs.(2) and (3) yield
$i\dot{a}_{n+1}=-\kappa_{eff}b_{n}$ and
$i\dot{b}_{n}=-\kappa_{eff}^{*}a_{n+1}$ with
$\kappa_{eff}=\kappa_{ab} \exp(i \phi)$ and $\phi=\sigma z+
\Theta$, i.e. waveguide B$_{n}$ turns out to be coupled solely
with waveguide A$_{n+1}$. For a propagation length $z_0$ equal to
one coupling length, light trapped in waveguide B$_n$ is thus
fully transferred into waveguide A$_{n+1}$. Therefore, after a
full modulation cycle, {\it an arbitrary} light distribution in
the primary array A$_n$ is shifted, without distortion, by one
 unit from the left to the right. The result is a net drift of
refracted light, with a locked refraction angle
$\theta_r=d/(2z_0)$ {\it independent of the initial incidence
angle $\theta$ and beam shape}, and a suppression of discrete
diffraction. Similarly, an arbitrary light distribution in the
secondary array B$_n$ is shifted by one unit but this time from
the right to the left. As the ratchet effect results from an
alternating synchronization of couples of waveguides in the array
($A_n \rightarrow B_n$ and $B_n \rightarrow A_{n+1}$), different
bending profiles $x_0(z)$, such as sequences of
sinuosoidally-curved waveguides with alternating amplitudes
mimicking the driving field of Ref.\cite{Creffield07}, do not
generally generate a ratchet effect because of incomplete
decoupling of the transport dynamics. As compared to other ratchet
schemes in tight-binding lattices
\cite{Gong07,Creffield07,Isart07}, our zigzag configuration offers
the advantage of spatially separating a primary lattice (in which
we want to realize a net transport) from an auxiliary one, which
is transiently excited.\\
The above analytical results are strictly valid within a
tight-binding model of Eq.(1) and in the averaging limit, which
neglects weak light transfer among waveguides due to non-resonant
coupling terms. We checked the possibility of observing
rectification of light refraction beyond such approximations by a
beam propagation analysis of Eq.(1). In the simulations, circular
channel waveguides with a super-Gaussian index profile $\Delta
n(x,y)=\Delta n_{A,B} \exp\{-[(x^2+y^2)/w^2]^3 \}$, with radius
$w$ and index changes $\Delta n_A$ and $\Delta n_B$ in the primary
and auxiliary arrays, are assumed. The vertical distance $d_v$
between the two arrays is chosen to be $d_v=\sqrt 3 d /2$, i.e.
the zigzag is defined by an equilateral triangle. As an input
beam, we typically assumed an elliptical Gaussian beam, with a
tilting angle $\theta$ (in the $x$ direction) smaller than the
Bragg angle $\theta_B$ to excite the lowest band of the array. The
vertical beam position and size are adjusted to couple a few
waveguides of the primary array. Examples of rectified light
refraction, observed in the curved arrays for a few values of beam
incidence angle $\theta$, are shown in Fig.2. In the figure, the
integrated beam intensity distribution $\int |\psi(x,y)|^2dy$
versus propagation distance $z$ is shown and compared to the
behavior that one would observe if the arrays were straight, i.e.
with usual discrete light refraction patterns. For the chosen
array parameters and wavelength, the coupling rates and
propagation mismatch are estimated to be $\kappa_{aa}=\kappa_{bb}
\simeq \kappa_{ab} \simeq 1.939 \; {\rm cm}^{-1}$ and $\sigma
\simeq 16.19  \; {\rm cm}^{-1}$, corresponding to a radius of
curvature $R=4$ cm and a semi-cycle period $z_0=8.1$ mm according
to Eq.(4). The numerical results clearly show that, for curved
arrays, light refraction always occurs at the angle
$\theta_r=d/(2z_0)$, regardless of the beam incidence angle
$\theta$, and discrete diffraction is well suppressed as a result
of the controlled transport mechanism. Of course beam refraction
is reversed when the position of the elliptical input beam is
shifted to excite waveguides of the auxiliary array. As an
example, Fig.3 shows rectification of light refraction when the
array B$_n$ is excited at the input by an elliptical Gaussian beam
at normal incidence. Note that, as compared to
Fig.2(a), the refraction angle of discretized light is now reversed.\\
In conclusion, rectification of light refraction in a periodic
photonic structure, which provides an example of an Hamiltonian
ratchet system, has been proposed, together with a possible
experimental implementation based on periodically-curved zigzag
waveguide arrays.

\begin{figure}[htb]
\centerline{\includegraphics[width=8.2cm]{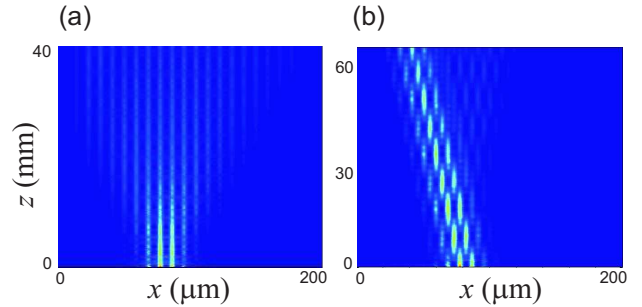}}
\caption{(Color online) Same as Fig.2(a), but for an elliptical
Guassian input beam vertically displaced to excite waveguides of
the auxiliary array. In (a) the arrays are straight, whereas in
(b) they are periodically-curved.}
\end{figure}

Author E-mail address: longhi@fisi.polimi.it


\end{document}